\documentclass[twocolumn,superscriptaddress,aps,prx]{revtex4-2}
\usepackage{graphicx}
\usepackage{dcolumn}
\usepackage{bm}
\usepackage{subfigure}
\usepackage{amsmath}
\usepackage{mathrsfs}
\usepackage{amssymb}
\usepackage{setspace}
\usepackage{array}
\usepackage{multirow}
\usepackage{float}
\usepackage{flushend}
\usepackage{footmisc}
\usepackage[pdfstartview=FitH,
CJKbookmarks=true,
colorlinks,
linkcolor=blue,
anchorcolor=blue,
citecolor=blue,
urlcolor=blue,
]{hyperref}

\usepackage{footmisc}

\renewcommand{\footnoterule}{
	\kern -4pt  
	\hrule width 0.18\linewidth height 0.6pt
	\kern 12pt 
}

\usepackage{footmisc}

\usepackage{braket}
\usepackage{braket}
\begin{document}

\global\long\def\id{\mathbbm{1}}
\global\long\def\ui{\mathbbm{i}}
\global\long\def\ud{\mathrm{d}}

\title{Quantum Mpemba Effect in Dissipative Spin Chains at Criticality}
\author{Zijun Wei}
\affiliation{School of Physics, Nankai University, Tianjin 300071, China}
\author{Mingdi Xu}
\affiliation{School of Physics, Nankai University, Tianjin 300071, China}

\author{Xiang-Ping Jiang}
\affiliation{School of Physics, Hangzhou Normal University, Hangzhou, Zhejiang 311121, China}

\author{Haiping Hu}
\email{hhu@iphy.ac.cn}
\affiliation{Beijing National Laboratory for Condensed Matter Physics, Institute of Physics, Chinese Academy of Sciences, Beijing 100190, China}
\affiliation{School of Physical Sciences, University of Chinese Academy of Sciences, Beijing 100049, China}

\author{Lei Pan}%
\email{panlei@nankai.edu.cn}
\affiliation{School of Physics, Nankai University, Tianjin 300071, China}

\begin{abstract}

The Quantum Mpemba Effect (QME) is the quantum counterpart of the classical Mpemba effect--a counterintuitive phenomenon in which a system initially at a higher temperature relax to thermal eauilibrium faster than one at a lower temperature. In this work, we investigate the QME in one-dimensional quantum spin chains coupled to a Markovian environment. By analyzing the full relaxation dynamics governed by the Lindblad master equation, we reveal the emergence of a strong quantum Mpemba effect at quantum critical points. Our findings reveal that criticality enhances the non-monotonic dependence of relaxation times on the initial temperature, leading to anomalously accelerated equilibration. This phenomenon is directly linked to the structure of the Liouvillian spectrum at criticality and the associated overlaps with the initial states. These findings demonstrate that quantum phase transitions could provide a natural setting for realizing and enhancing non-equilibrium phenomena in open quantum systems.
\end{abstract}

\maketitle

\section{Introduction}
The Mpemba effect~\cite{ME}, originally reported by the Tanzanian student Erasto Mpemba, refers to the counterintuitive observation that, under the same environmental conditions, a system starting from a higher temperature may cool more rapidly than one beginning at a lower temperature. Despite early controversy surrounding the Mpemba effect, it has been documented across a range of classical experimental setups~\cite{ME_classical1,ME_classical2,ME_classical3,ME_classical4,ME_classical5,ME_classical6,ME_classical7,ME_classical8,ME_classical9}. In parallel, theoretical analyses have suggested the possibility of an inverted version, where cooler systems heat faster than warmer ones, a prediction that has recently been validated in laboratory experiments~\cite{Inverse_ME1,Inverse_ME2,Inverse_ME3}.
While traditionally considered a classical anomaly, recent theoretical and experimental studies have extended the concept to quantum domain, giving rise to the so-called quantum Mpemba effect (QME) where a quantum system initialized further from equilibrium may relax faster toward a steady state than one closer to equilibrium. 

The QME has attracted increasing attention following its realization in engineered quantum platforms~\cite{QME_Exp1,QME_Exp2,QME_Exp3}. These observations have led to intensive theoretical efforts to understand the underlying mechanisms in various physical regimes, including quantum integrable systems~\cite{QME1,QME2,QME3,QME4,QME41}, disordered systems~\cite{QME5,QME6,QME7}, random quantum circuits~\cite{QME8,QME9,QME901}, and other settings ~\cite{QME10,QME101,QME105,QME11,QME12,QME13,QME14,QME15,QME16,QME17,QME18,QME19,QME20,QME21}.
A natural theoretical framework for describing QME and its inverse is through open quantum systems. The experimental realization of open quantum systems in a variety of controllable platforms \cite{Exp1,Exp2,Exp3,Exp4,Exp5,Exp6,Exp7,Exp8,Exp_new1,Exp_new2,Exp_new3} has reinvigorated research into dissipative quantum dynamics governed by master equations. They provide means to characterize QME by analyzing relaxation pathways from different initial states and investigating how decoherence shapes the approach to steady state.
In open quantum systems governed by Lindblad dynamics, the QME manifests as a non-monotonic dependence of relaxation times on initial conditions. This behavior is rooted in the spectral structure of the Liouvillian superoperator. Depending on how the initial state overlaps with slow-decaying modes, relaxation can be anomalously accelerated.

\begin{figure}
	\centering
	\includegraphics[width=1\linewidth]{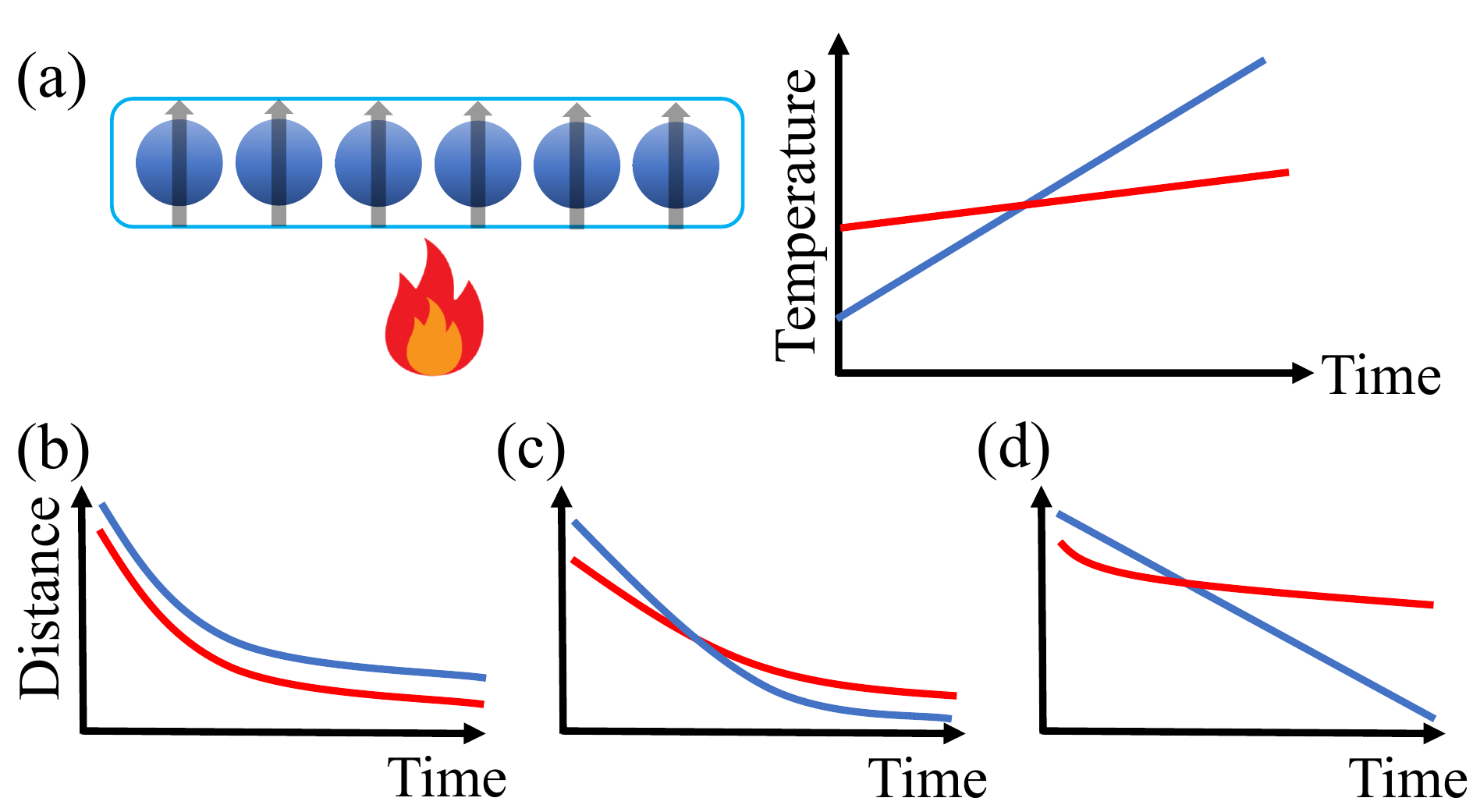}
	\caption{Schematic illustration of the quantum Mpemba effect, where "distance" denotes the deviation from the thermal equilibrium state. (a) Illustration of the inverse Mpemba effect. (b) Conventional relaxation dynamics with no Mpemba effect-no trajectory crossing occurs. (c) Weak Mpemba effect (wQME) characterized by trajectory crossing that depends on initial conditions. (d) Strong Mpemba effect (sQME), where trajectory crossing always occurs regardless of the initial distance from equilibrium.}
	\label{fig0}
\end{figure}

In this work, we explore the emergence of the QME in one-dimensional quantum spin chains near quantum criticality, focusing on the XXZ and $J_1-J_2$ XXZ models subject to Markovian bath. 
Specifically, we consider dephasing-type dissipative operators, for which the system is eventually heating to the maximally mixed state, corresponding to the infinite-temperature limit, regardless of whether the initial state is effectively hotter or colder. Under these dynamics, two thermal states with different initial temperatures may exhibit distinct heating rates, as illustrated in Fig.~\ref{fig0}(a). Generally, the heating process from two thermal states at different initial temperatures can be classified into three distinct regimes, depending on their relative distances from equilibrium and their overlaps with the slow Liouvillian decay modes. The first is the absence of QME, where both the initially close and initially far states follow similar relaxation trajectories without crossing, as shown in Fig.~\ref{fig0}(b). The second is the weak QME (wQME), where trajectory crossings occur but are sensitive to the temperature difference; the effect may disappear when the initial distance between the states becomes large, as shown in Fig.~\ref{fig0}(c). The third is the strong QME (sQME) \cite{QME_Exp3}, which is temperature-independent and persists regardless of the initial separation between the states and trajectories always cross, as illustrated in Fig.~\ref{fig0}(d).

We show that the QME becomes markedly pronounced at critical points, where the system exhibits a sQME, characterized by regimes in which states prepared farther from equilibrium relax more rapidly than all those initialized closer to the steady state. Unlike the wQME, which only requires a non-monotonic dependence of the relaxation time on the initial condition, the strong form implies a global minimum of the relaxation time for a particular far-from-equilibrium initial state. This behavior arises when such a state has minimal overlap with those slow decaying modes of the Liouvillian, leading to anomalously fast relaxation. Away from criticality, the sQME degrades into a wQME and then disappears entirely. Our analysis reveals that criticality strongly enhances the separation of Liouvillian decay modes and leads to sharp changes in relaxation behavior. We demonstrate that this enhancement is tightly localized near the critical anisotropy parameter (e.g., $\Delta = 1$ in the XXZ chain) and becomes more sensitive with increasing system size and temperature. These findings are further supported by extending our analysis to the $J_1-J_2$ XXZ model, where the presence of competing interactions gives rise to an additional quantum phase transition. In both models, we find that quantum critical points act as natural amplifiers of the QME, offering a new perspective on the interplay between criticality and dissipation in non-equilibrium quantum many-body dynamics.

The rest of this paper is organized as follows. In Sec.~\ref{QME_XXZ}, we introduce the theoretical framework of open quantum systems and the model considered in this work.  Sec.~\ref{sQME} discusses the dissipative dynamics in spin chain models, and present our main numerical results, including the characterization of QME and the identification of sQME at quantum critical points. In Section~\ref{Theory_Experiment}, we analyze the emergence of sQME from overlaps between initial states and slow modes, and also the potential realization in controllable quantum simulators. Finally, Sec.~\ref{Conclusion} summarizes our findings and outlines possible directions for future research.


\section{Theoretical Framework and Model}
\label{QME_XXZ}
A natural framework for exploring the QME is provided by the dissipative time evolution of open quantum systems.
Specifically, we consider a quantum system coupled to an external reservoir, where the total Hamiltonian is given by
\begin{align}
	H_{T} = H_{S} + H_{R} + H_{SR},
\end{align}
where $H_{S}$ and $H_{R}$ describe the system and reservoir, and $H_{SE}$ represents their coupling term.  
Within the Born-Markov approximation~\cite{Moy1999,Breuer2002}, eliminating the  reservoir's degrees of freedom yields the Lindblad master equation~\cite{Lindblad1,Lindblad2}
\begin{align}
	\frac{d\rho(t)}{dt} = \mathscr{L}[\rho(t)] = -i[H_{S}, \rho(t)] + \mathcal{D}[\rho(t)], \label{Lindblad_Eq}
\end{align}
where $\rho(t)$ denotes the density matrix of the system, and $\mathscr{L}$ represents the Liouvillian superoperator, which ensures the preservation of both trace and positivity.
The commutator term captures coherent unitary evolution, whereas the dissipator $\mathcal{D}[\rho(t)]$ incorporates environmental dissipation effects with the following form
\begin{align}
	\mathcal{D}[\rho(t)] = \sum_{j} \gamma_{j} \left( L_{j} \rho L^{\dagger}_{j} - \frac{1}{2} \left\{ L^{\dagger}_{j} L_{j}, \rho \right\} \right).
\end{align}
Here $\{\cdot,\cdot\}$ denotes the anticommutator, $L_{j}$ are the quantum jump (dissipation) operators, $j$ labels the lattice sites, and $\gamma_{j}$ is the corresponding dissipation strength.

Analogous to how the Hamiltonian dictates the evolution of the Schr{\"o}dinger equation in closed systems, the Liouvillian spectrum governs the entire non-equilibrium dynamics of open systems, with the formal solution given by $\rho(t) = e^{\mathscr{L}t}[\rho_{0}]$. In the long-time limit, the system relaxes to a steady state $\rho_{\mathrm{ss}} = \lim_{t\to\infty} \rho(t)$, which corresponds to the right eigenvector of $\mathscr{L}$ associated with the zero eigenvalue.
More generally, the solution admits a spectral decomposition,
\begin{align}
	\rho(t) = \rho_{\mathrm{ss}} + \sum_{n=2}^{d^{2}} \mathrm{Tr}(l_{n}\rho_{0})\, r_{n} \, e^{\lambda_{n} t}, \label{rho_extend}
\end{align}
where $d$ is dimension of the Hilbert space, $\rho_{0}$ is the initial state, and $\{r_{n}\}$ and $\{l_{n}\}$ are, respectively, the right and left eigenvectors of $\mathscr{L}$ with eigenvalues $\lambda_{n}$.  
These satisfy $\mathscr{L}[r_{n}] = \lambda_{n} r_{n}$ and $\mathscr{L}^{\dagger}[l_{n}] = \lambda_{n}^{*} l_{n}$, with eigenvalues ordered according to their real parts as $0=\lambda_{1} \le |\mathrm{Re}(\lambda_{2})| \le |\mathrm{Re}(\lambda_{3})| \le \dots \le |\mathrm{Re}(\lambda_{d^{2}})|$, all having non-positive real components.

Equation~\eqref{rho_extend} shows that the relaxation process can be decomposed into $d^{2}$ distinct decay channels.  
The mode $\lambda_{1}$ represents the stationary solution, while all other modes relax exponentially with rates determined by $|\mathrm{Re}(\lambda_{n})|$.  
The smallest nonzero value, $|\mathrm{Re}(\lambda_{2})|$, sets the slowest decay scale and thus controls the leading subdominant contribution to the late-time dynamics. It is straightforward to see that the smaller the overlap with slow modes of a given state, the faster it relaxes, and vice versa.
Therefore, comparing the weight distributions of different states on the slow modes provides a crucial criterion for identifying the presence of a QME in the dynamics. 

To explore the quantum QME in open systems, we consider the following type of XXZ spin chain with the Hamiltonian 

\begin{equation}\label{Hamiltonian}
	H=\sum_{n=1}^2\sum_j J_n\left(S_i^xS_{j+n}^x+S_j^yS_{j+n}^y+\Delta_n S_j^zS_{i+n}^z\right),
\end{equation}
where $S_n^{x,y,z}$ are spin-1/2 operators on site $j$, $J_{1}$ and $J_{2}$ denote the exchange coupling strengths between nearest-neighbor and next-nearest-neighbor lattice sites, respectively, and $\Delta_1$, $\Delta_2$ are the anisotropy parameters.
The case $J_{1}=-J, J_{2} = 0$, $\Delta_2=0,\Delta_1=\Delta$ corresponds to the XXZ model, whose ferromagnetic (antiferromagnetic) critical point is located at $\Delta = 1$ ($\Delta = -1$). 
When $J_{1} \neq 0$ and $J_{2} \neq 0$, the system realizes the $J_{1}$-$J_{2}$ XXZ model, with a critical point at $\Delta = 1$ for $J_{2}/J_{1} > -0.25$.

\section{Strong quantum Mpemba effect in at critical points}
\label{sQME}
\subsection{sQME in the XXZ model}
In the following, we first investigate the quantum Mpemba effect in the XXZ model (set to $J_2=0$, $J_1=-J$ and $\Delta_1=\Delta$ throughout). The model exhibits two quantum phase transitions as the anisotropy parameter $\Delta$ is varied. At $\Delta=1$ ($\Delta=-1$), the model undergoes a paramagnetic-antiferromagnetic (paramagnetic-ferromagnetic) transition exhibiting SU(2) symmetry. 
In the context of open quantum systems, the interaction between the system and its environment induces non-unitary dynamics, being  described by the Lindblad equation \eqref{Lindblad_Eq}. We consider the dissipation operator $L_j$ as dephasing, namely
\begin{equation}\label{eq3}
	L_j=S^z_j+\mathbb{I}/2,
\end{equation}
with $\mathbb{I}$ being the  $2\times 2$ identity matrix. After including the dissipative operator, the total magnetization along the $z$ direction remains conserved, making it natural and efficient to restrict the analysis to the invariant subspace with fixed $m=\langle S^z\rangle$.
Under the influence of environmental dephasing, the system eventually relaxes to the maximally mixed state, i.e., the infinite-temperature state. 
Therefore, based on the above properties, by preparing two initial states at different effective temperatures, we examine their heating dynamics in order to explore the possible emergence of the QME and to uncover its underlying physical mechanism. As initial states, we consider both the zero-temperature ground state and finite-temperature thermal states. Denote eigenstate basis of the Hamiltonian by $\{|\psi_{i}\rangle\}$, ordered by increasing eigenvalues, such that $|\psi_{0}\rangle$ corresponds to the ground state. The corresponding initial density matrix with temperature $T$ is defined as 
\begin{equation}\label{rho_T}
	\quad\rho_T=\sum_n\frac{e^{-\beta E_n}}{Z}|\psi_n\rangle\langle\psi_n|,
\end{equation}
where $\beta=1/(k_BT)$, with $k_B$ representing the Boltzmann constant and $T$ the temperature of the thermal state. Throughout this work, we set the Boltzmann constant $k_{B}=1$. $Z=\sum_ie^{-\beta E_i}$ is the partition function ensuring normalization. Intuitively, finite-temperature ($T>0$) initial states are closer to the steady-state than the zero-temperature ground state, in terms of their ``distance" to the maximally mixed state.
To quantify the distance between the evolving density matrix $\rho(t)$ and the steady-state $\hat{\rho}_{ss}$, we utilize the squared $L^2$-norm defined by 
\begin{equation} 
	D(t)=\|\rho(t)-\rho_{\mathrm{ss}}\|_{L^2}^2=\mathrm{Tr}[(\rho(t)-\rho_{\mathrm{ss}})^2].\label{Distance}
\end{equation}
The $L^2$-norm provides a straightforward and computationally efficient measure of the ``distance" or distinguishability between two quantum states, making it a practical tool for tracking relaxation dynamics.
It then follows that if the zero-temperature initial state, which is farther from the steady state, relaxes faster than a finite-temperature initial state that is closer to the steady state, the QME occurs. 
\begin{figure}
	\centering
		\includegraphics[width=1.0\linewidth]{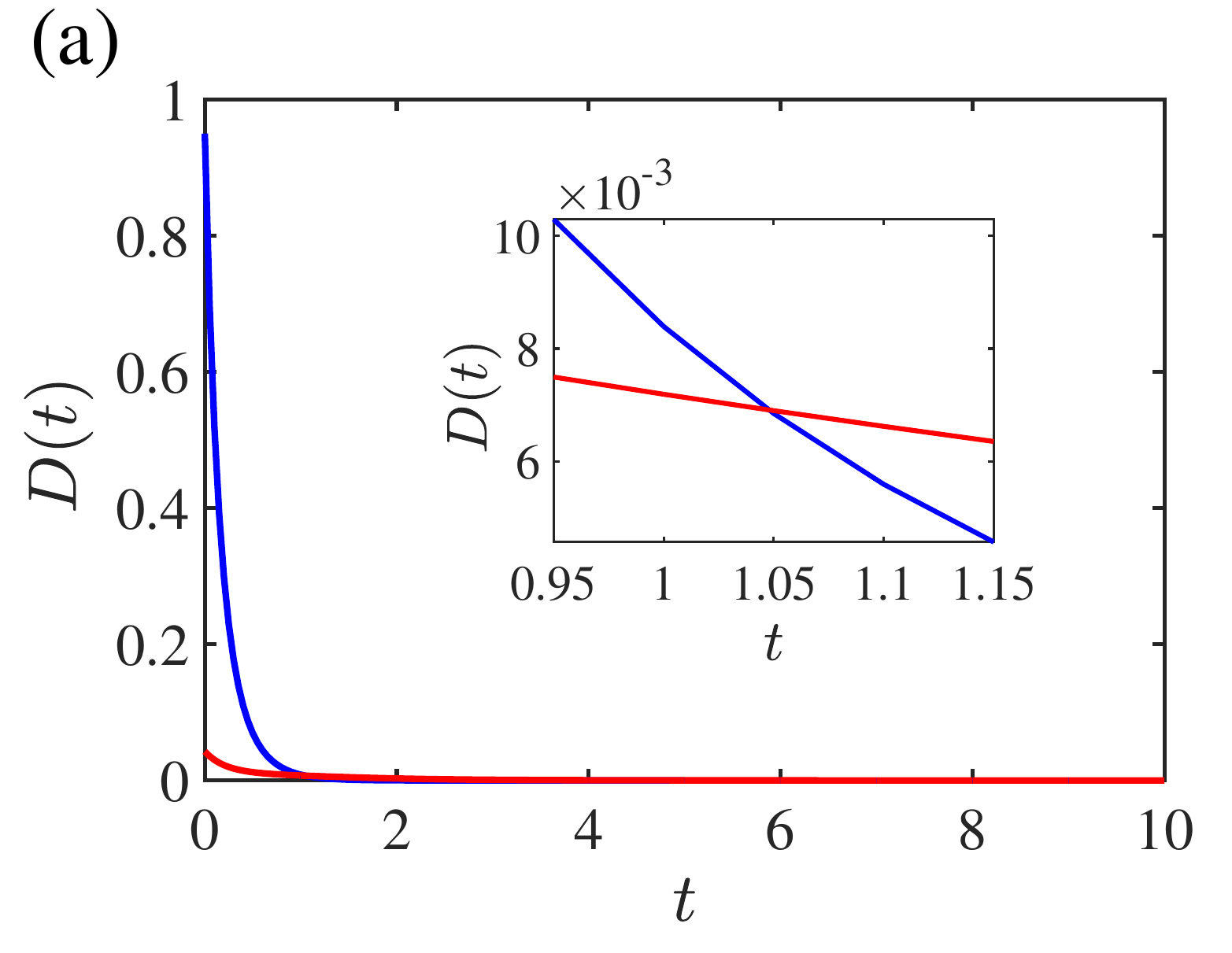}
	\includegraphics[width=1.0\linewidth]{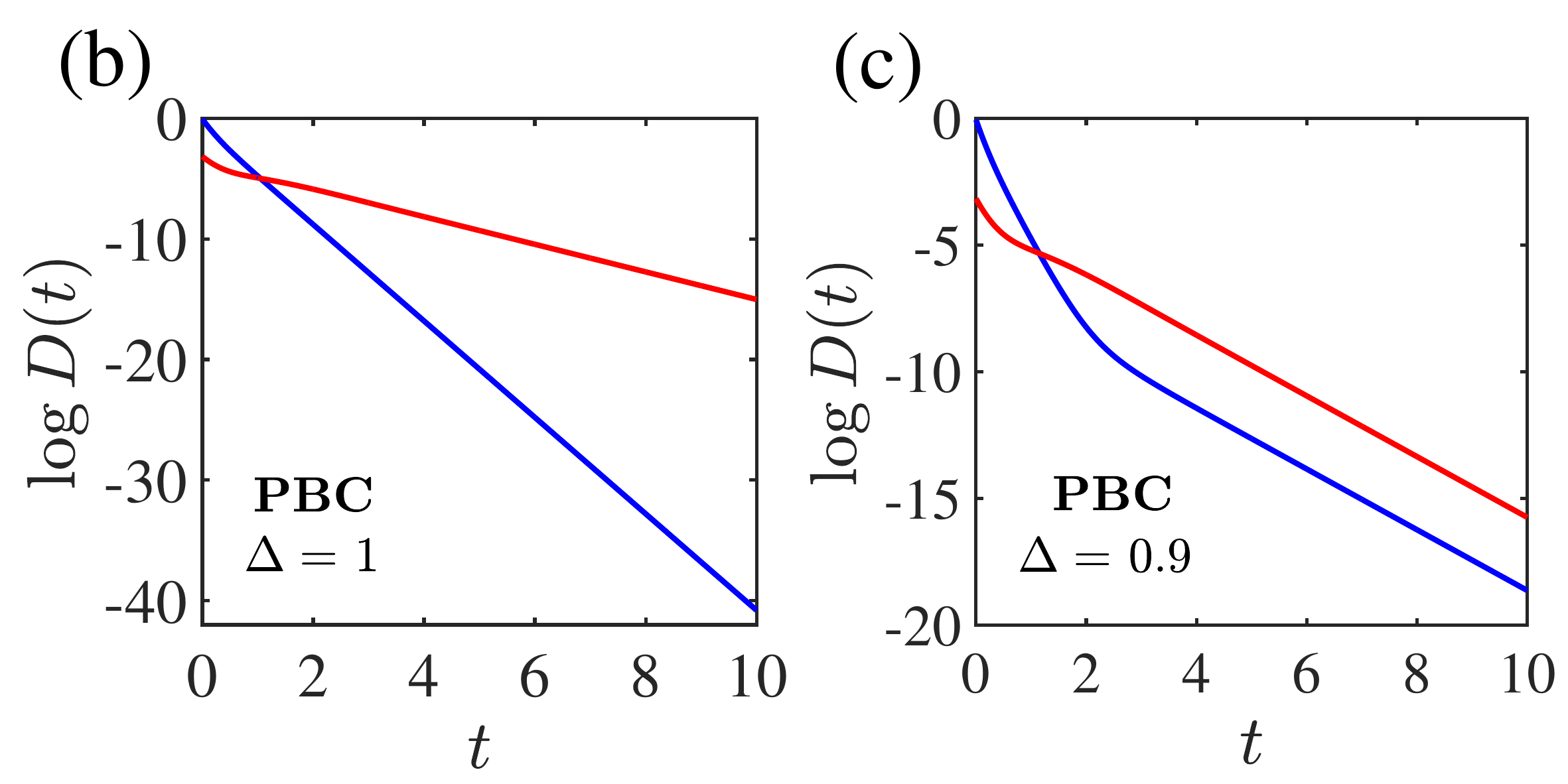}
		\includegraphics[width=1.0\linewidth]{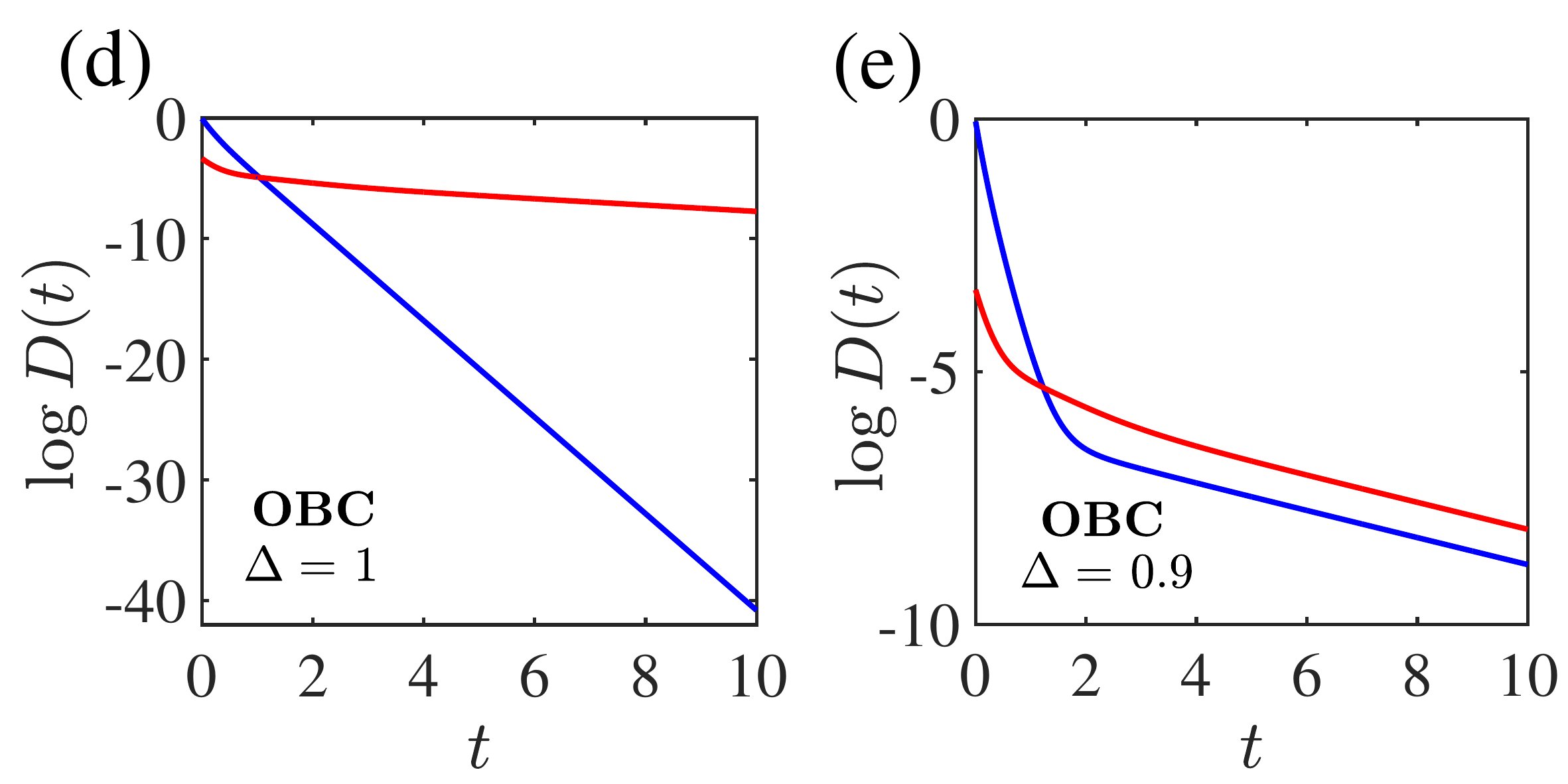}
	\caption{Dynamical evolution in the XXZ model.  (a) Time evolution of the distance $D(t)$ at ferromagnetic transition point $\Delta=1$ for two different initial states. The inset shows a magnified view where a crossing occurs between the evolutions of different initial states, indicating the emergence of the QME. (b-c) Under periodic boundary conditions, we observe a sQME at $\Delta=1$, while for $\Delta \ne 1$, only wQME appears. (d-e) Similar behavior is observed under open boundary conditions: a sQME occurs at $\Delta=1$, whereas deviations from this point exhibit only wQME. Here we set the system size $L=8$, total spin $S_{\mathrm{tot}}^z=0$.  The blue line denotes time evolution from zero-temperature initial and the red one is set to the finite temperature with $T/J=1$.}
	\label{fig1}
\end{figure}

We first discuss the QME in the XXZ model  ($J_{2}=0$). The system evolves from two initial states with different temperatures under the Lindblad master equation~\eqref{Lindblad_Eq}, and the results are plotted in Fig.~\ref{fig1}. 
As illustrated in Fig.~\ref{fig1}(a), at paramagnetic-ferromagnetic transition point ($\Delta=1$), the zero-temperature state (blue curve) starts farther from the steady state which is the identity matrix corresponding to an effective infinite-temperature state than the finite-temperature state (red curve) at the initial time. 
Nevertheless, the two trajectories cross during the evolution, demonstrating that the zero-temperature state actually approaches the steady state faster than the finite-temperature one in the heating process. 
This crossing behavior serves as a clear signature of the QME. 

Figures~\ref{fig1}(b-e) further reveal the role of the anisotropy parameter $\Delta$. 
At $\Delta=1$, we observe a sQME, where the trajectory of zero-temperature initial state intersects with those of all $T>0$ initial states, providing direct evidence for the presence of the sQME. 
To highlight the critical nature of sQME, we calculate the dynamics at a slight deviation from $\Delta=1$. One can see immediately that the sQME is reduced to a wQME, meaning that the crossing behavior between the zero-temperature state and finite-temperature states no longer occurs for all $T>0$, but only within a limited range of temperatures. 
In other words, the occurrence of QME becomes dependent on the choice of the finite-temperature initial state, reflecting its weaker and more restricted nature.
We further investigate the influence of boundary conditions on the QME. We discuss the impact of boundary conditions by contrasting periodic boundary conditions (PBC, Figs.~\ref{fig1}(b,c)) with open boundary conditions (OBC, Figs.~\ref{fig1}(d,e)). While PBC tends to accelerate the relaxation of both initial states, the qualitative features of the QME whose presence and relative strength remain essentially unchanged, especially near the critical point $\Delta=1$.

\begin{figure}
	\centering
	\includegraphics[width=1\linewidth]{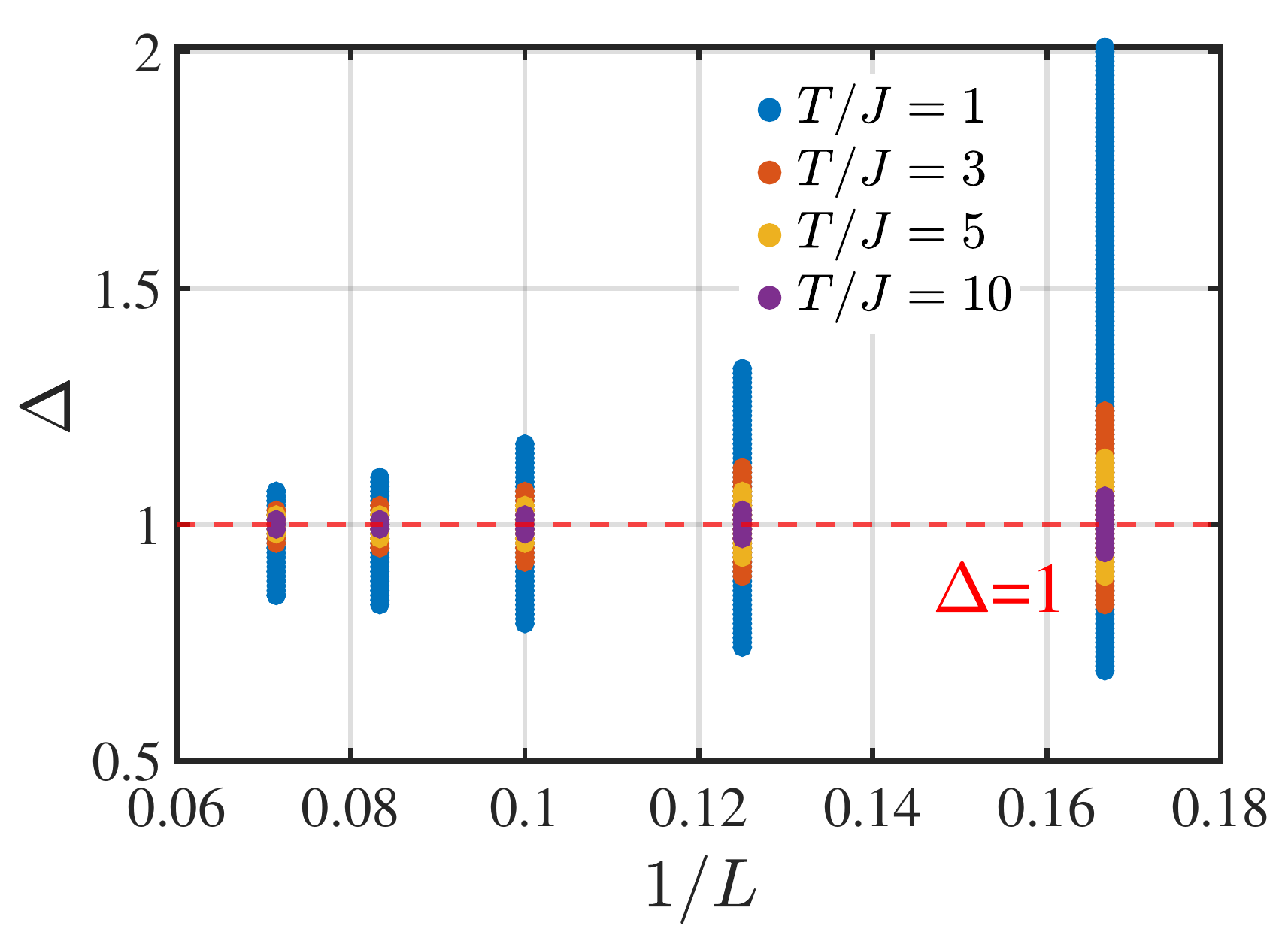}
	\caption{The parameter range of anisotropy parameter $\Delta$ supporting the emergence of the QME as a function of system size, with total magnetization fixed at zero. As the initial temperature increases and the system size grows, the range of $\Delta$ values exhibiting QME becomes progressively narrower, eventually collapsing to a region in the vicinity of the quantum critical point. The anisotropy parameter $\Delta$ is scanned from $0$ to $3$ in steps of $0.01$. }
	\label{fig2}
\end{figure}

Furthermore, we perform a systematic analysis of the conditions under which the QME emerges. Specifically, we vary the system size ($L=6,8,10,12,14$), the temperature of finite-temperature initial states, and the anisotropy parameter $\Delta$.
The results are summarized in Fig.~\ref{fig2} in which we find that signatures of the QME appear exclusively in the vicinity of the paramagnetic-ferromagnetic quantum critical point at $\Delta=1$, while they are completely absent away from criticality. 
In addition, increasing the system size or raising the initial temperature systematically reduces the parameter window in which the QME is observed. 
This shrinking region of occurrence indicates that finite-size and finite-temperature effects are essential for the manifestation of the QME. 
Taken together, these observations provide compelling evidence that in the thermodynamic limit $L\to\infty$, the sQME can persist only exactly at the quantum critical point.

\begin{figure*}[ht!]
	\centering
	\includegraphics[width=1\linewidth]{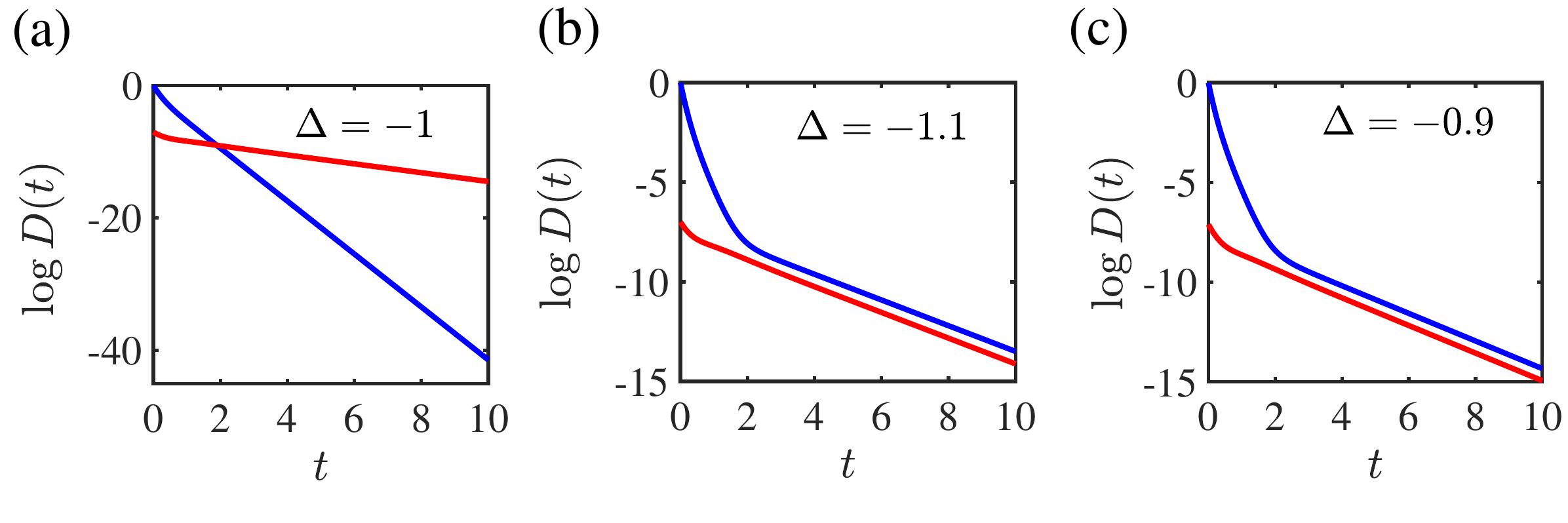}
	\caption{Dynamical evolution near the antiferromagnetic quantum critical point of the XXZ model. The initial states are chosen as the highest excited state (blue line) and a finite negative-temperature state (red line), with system size $L=8$. (a) At the critical point $\Delta=-1$, sQME emerges. (b) For $\Delta=-1.1$, no QME is observed. (c) For $\Delta=-0.9$, no QME is observed. The blue line denotes the evolution from initial state with temperature $T=0^-$ and the red one is $T=-5$.}
	\label{fig5}
\end{figure*}

Since the XXZ model also hosts an antiferromagnetic quantum critical point, one might naturally expect the emergence of a sQME there as well. 
This can in fact be understood from a simple symmetry consideration. 
The antiferromagnetic XXZ Hamiltonian differs from the ferromagnetic one only by an overall sign change ($J \to -J$), implying that the two models share an identical spectrum, with the only difference being that positive (negative) energy eigenstates of the ferromagnetic Hamiltonian correspond to negative (positive) energy eigenstates of the antiferromagnetic Hamiltonian with opposite ordering. 
As a consequence, the ground state of the ferromagnetic Hamiltonian, which serves as the zero-temperature state in our previous analysis, corresponds to the highest-energy eigenstate of the antiferromagnetic Hamiltonian and thus to an effective state at negative zero temperature $T=0^-$. 
More generally, a finite-temperature state ($T>0$) of the ferromagnetic Hamiltonian maps onto a thermal state of the antiferromagnetic Hamiltonian at negative temperature $-T$. 
This mapping follows directly from Eq.~\eqref{rho_T}, since flipping the sign of all energies is equivalent to replacing the temperature parameter by its negative. As shown in Fig.~\ref{fig5}(a), similar to the ferromagnetic critical point, numerical results demonstrate the existence of a sQME at the antiferromagnetic critical point ($\Delta=-1$). However, as illustrated in Figs.~\ref{fig5}(b) and \ref{fig5}(c), the sQME vanishes once the system deviates from this critical point.

\subsection{sQME in the $J_1-J_2$ XXZ model}

Having established the occurrence of sQME at both the ferromagnetic ($\Delta=1$) and antiferromagnetic ($\Delta=-1$) critical points of the standard XXZ chain, we now turn to a more generalized setting, namely the $J_{1}$–$J_{2}$ XXZ model. This model introduces next-nearest-neighbor interactions in addition to the conventional nearest-neighbor coupling, thereby enriching the phase diagram with competing orders and frustration effects. In what follows, we investigate whether sQME can also emerge in this extended model.

Previous numerical studies have demonstrated that the $J_{1}$-$J_{2}$ XXZ model exhibits a quantum phase transition when the anisotropy parameters take the isotropic values $\Delta_{1}=\Delta_{2}=\Delta=1$. 
In particular, when $J_{1}<0$ and $J_{2}>0$, the nearest-neighbor interactions promote ferromagnetic alignment, whereas the next-nearest-neighbor interactions favor anti-ferromagnetic ordering. This competition between ferromagnetic and anti-ferromagnetic tendencies gives rise to a nontrivial critical coupling ratio that signals a quantum phase transition in the $J_{1}$-$J_{2}$ XXZ model. 
The $J_{1}$-$J_{2}$ chain hosts a ferromagnetic phase for small $J_{2}/J_{1}$, and undergoes a transition into a gapped dimerized phase at the critical ratio $J_{2}/J_{1}=-0.25$ for the isotropic case $\Delta_{1}=\Delta_{2}=\Delta=1$ \cite{J1J2_1,J1J2_2}. Beyond this point, the next-nearest-neighbor anti-ferromagnetic interaction becomes strong enough to destabilize the ferromagnetic order, giving rise to a qualitatively different ground state structure.  Motivated by our above findings in the standard XXZ chain, it is natural to ask whether an analogous sQME may also emerge at this critical point of the $J_{1}$-$J_{2}$ model.

\begin{figure}
	\centering
	\includegraphics[width=1.0\linewidth]{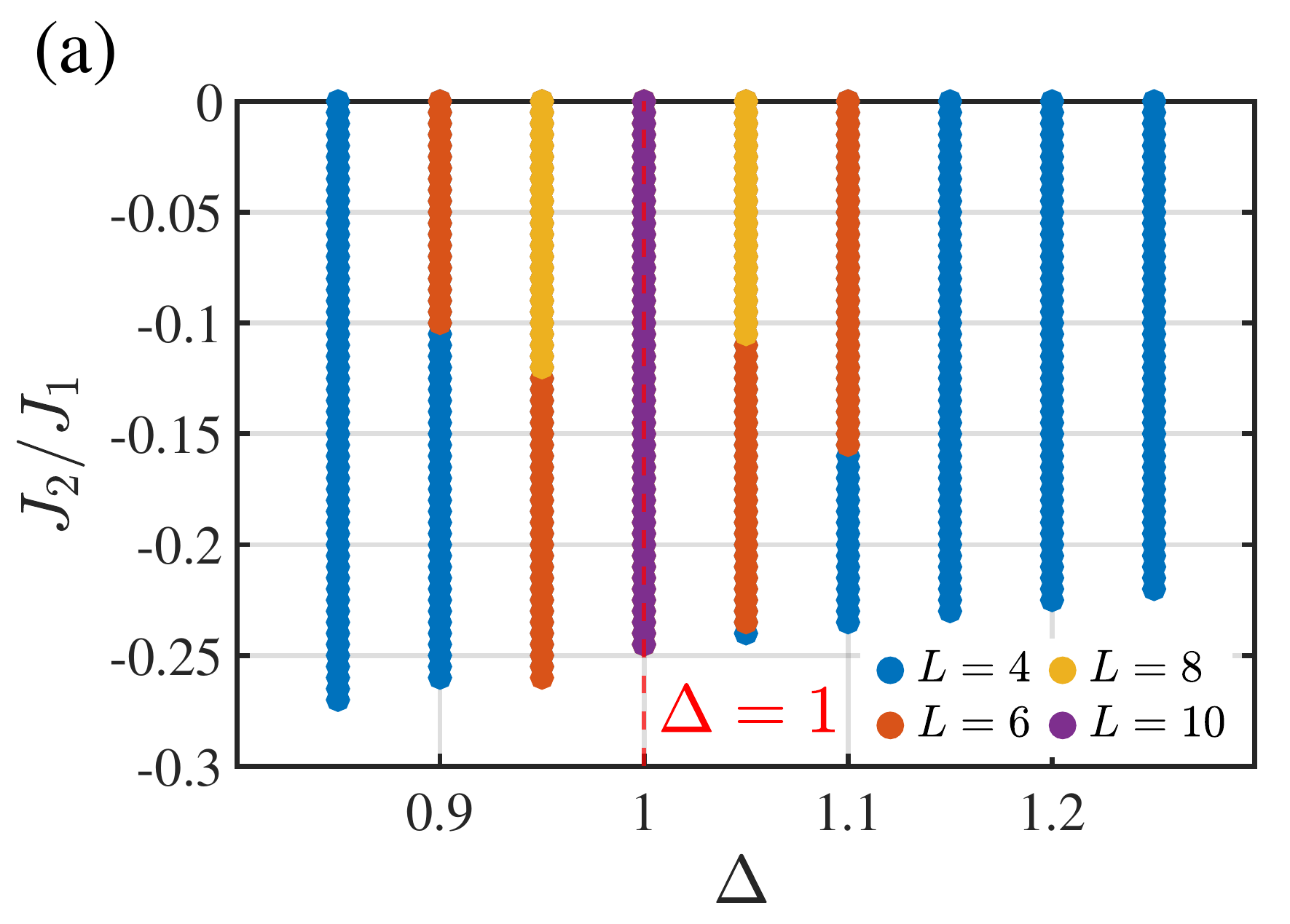}
	\includegraphics[width=1.0\linewidth]{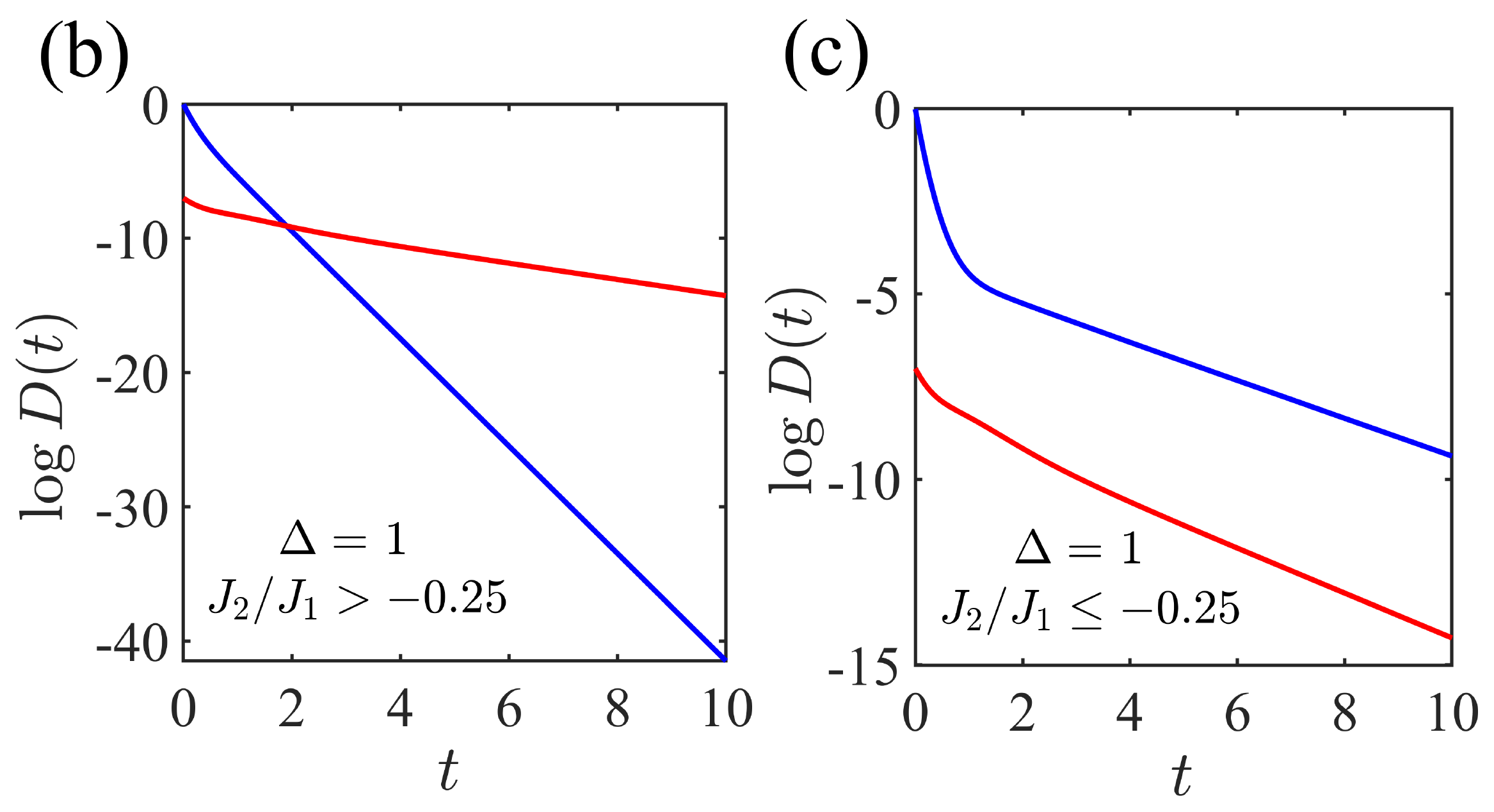}
	\caption{QME in the \( J_1-J_2 \) XXZ model, \( J_1 = 1 \), \( L = 8 \), \( T/J=1 \) and total spin is zero.  (a) It can be seen that as the system size increases, the parameter regime with $J_{2}/J_{1}<0$ exhibiting QME gradually shrinks. This suggests that in the thermodynamic limit ($L \to \infty$), QME persists only at $\Delta = 1$. (b) A sQME is observed at the quantum critical point \( J_2/J_1 > -0.25 \), \( \Delta = 1 \) where we set $J2/J1=-0.2499$. (c) When either the coupling ratio \( J_1/J_2 \) exceeds the critical threshold (\( J_2/J_1 \leq -0.25 \)), the sQME disappears entirely.}
	\label{fig3}
\end{figure}

A straightforward calculation shows that the system quickly departs from the sQME regime once either $\Delta_{1}$ or $\Delta_{2}$ deviates from unity. For this reason, our analysis below is restricted to the isotropic case $\Delta_{1}=\Delta_{2}=\Delta=1$ and we focus on the regime with $J_{1}<0$ and $J_{2}/J_{1}<0$. Our numerical simulations demonstrate that the $J_{1}$-$J_{2}$ XXZ chain supports a pronounced sQME when $J_{2}/J_{1}>-0.25$ at $\Delta=1$. Once the ratio decreases below this threshold, i.e., $J_{2}/J_{1}\leq-0.25$, the sQME collapses abruptly, as shown in Figs.~\ref{fig3}(b-c). This highlights the remarkable sensitivity of the effect to the critical point $J_{2}/J_{1}=-0.25$. In analogy with the standard XXZ model, the sQME in the $J_{1}$-$J_{2}$ chain survives exclusively along the isotropic line $\Delta=1$. Any deviation from this condition inevitably reduces it to a wQME or none QME. Moreover, increasing either the system size or the initial temperature further suppresses the parameter space where QME is visible, which eventually shrinks to the critical line $\Delta=1$, as illustrated in Fig.~\ref{fig3}(a).
Hence, our analysis demonstrates that the $J_{2}$-$J_{1}$ XXZ chain provides a second concrete example, beyond the standard XXZ model, where a sQME emerges precisely at the quantum critical point. The abrupt suppression of sQME once the system deviates from this transition highlights the intimate connection between nonequilibrium relaxation dynamics and underlying quantum phase transitions, reinforcing the view that criticality plays a decisive role in stabilizing sQME.

\section{Theoretical Explanation and Experimental realization}
\label{Theory_Experiment}

In this section, we provide a theoretical explanation for the emergence of the sQME and discuss its possible experimental realization and detection. 
The mechanism can be understood in terms of the spectral properties of the Liouvillian superoperator, which governs the dissipative evolution. As shown in Eq.~\eqref{rho_extend}, the relaxation dynamics of a given initial state are dictated by the Liouvillian spectrum, with the relaxation times determined by the inverse of the real parts of the Liouvillian eigenvalues.
At late times, the relaxation dynamics can be approximated as  
\begin{equation}\label{time}
	\rho(t)-\rho_{ss} \sim e^{\mathrm{Re}(\lambda_{\text{max}})t} 
	\left[ \mathrm{Tr}\!\left( l_{\text{max}} \rho_0 \right) r_{\text{max}} e^{i\,\mathrm{Im}(\lambda_{\text{max}})t} \right],
\end{equation}
where $\lambda_{\text{max}}$ denotes the Liouvillian eigenvalue with the largest real part that has a nonzero overlap with the initial state $\rho_0$. 
The real part of $\lambda_{\text{max}}$ sets the asymptotic decay rate of the dynamics at long times. This implies that a larger in magnitude (i.e., a more negative) value of $\mathrm{Re}(\lambda_{\text{max}})$ corresponds to a faster relaxation of the system toward its steady state.
If an initial state has significant overlaps with the slow modes, the relaxation dynamics will be dominated by them, leading to a slower decay. In contrast, when the overlap with slow modes is small, the system predominantly relaxes through fast modes, resulting in a faster decay. From this perspective, the emergence of the sQME can be attributed to the fact that zero-temperature initial state has a much smaller overlap with the slow modes of the Liouvillian spectrum compared to finite-temperature initial states. Since these modes govern long relaxation times, such overlap suppresses slow relaxation channels and effectively accelerates the overall approach to equilibrium compared to states prepared at different initial temperatures.

\begin{figure}[ht!]
	\centering
	\includegraphics[width=1\linewidth]{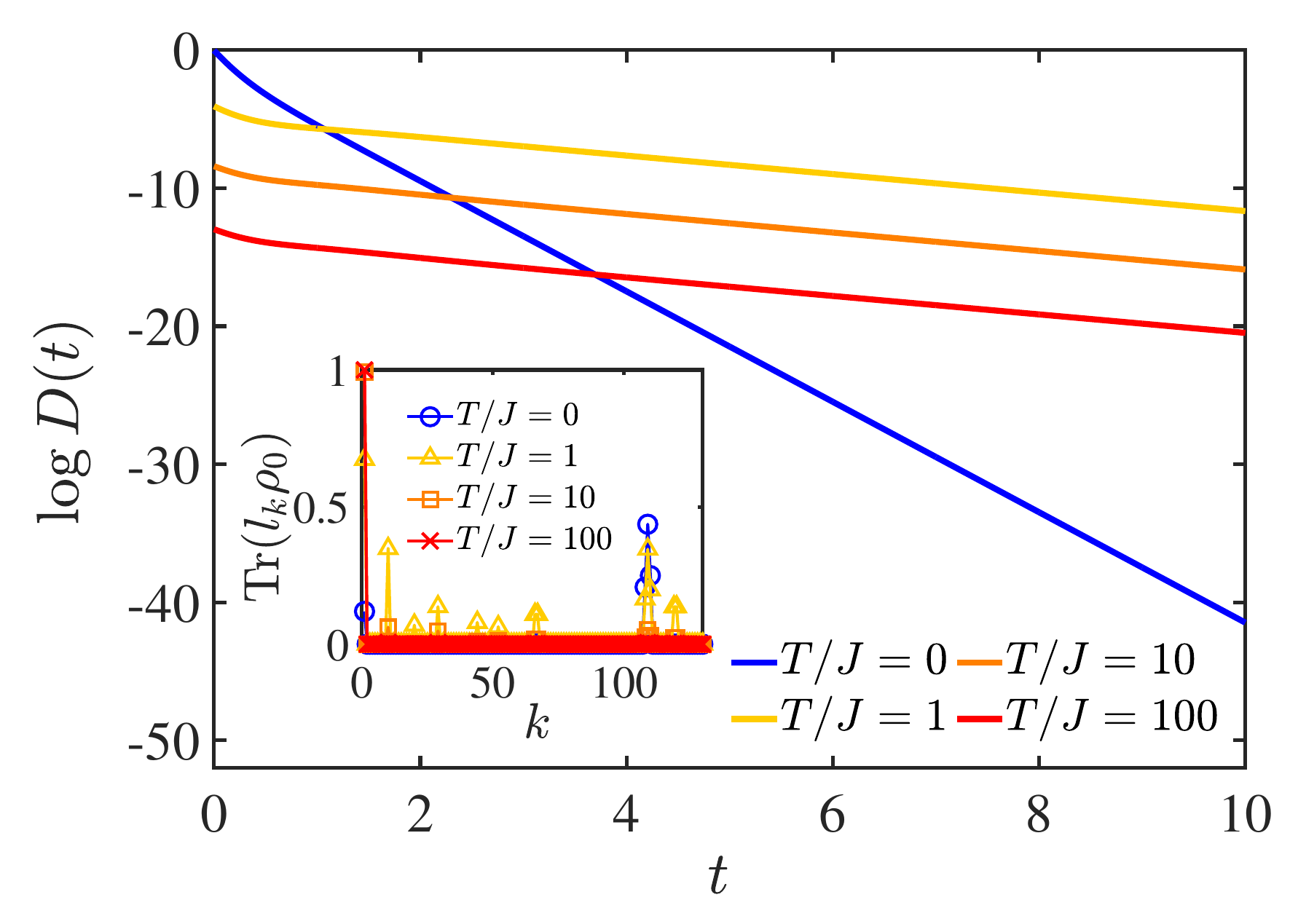} 
	\caption{Dissipative dynamics of the XXZ model at the ferromagnetic critical point ($\Delta=1$) for the zero-temperature ground state and finite-temperature initial states  ($ T=1J,\, T=10J,\, T=100J$). The main panel shows the heating dynamics, where the zero-temperature trajectory crosses all finite-temperature trajectories, signaling the presence of the sQME. The inset displays the overlap distributions of the initial states with the Liouvillian eigenmodes , where the vertical axis denotes the overlap magnitude and the horizontal axis represents the eigenmode index (ordered by the real part of the Liouvillian eigenvalues from large to small). For finite-temperature states, the overlaps include significant weight on the slow modes, leading to intrinsically slow relaxation, whereas the zero-temperature state has vanishing overlap with slow modes and instead couples to modes with large negative real parts ($k>100$). }
	\label{fig6}
\end{figure}

Based on the above analysis, we now explicitly calculate the overlap distribution of different initial states, namely the zero-temperature ground state and finite-temperature thermal states, with the Liouvillian eigenmodes. Figure~\ref{fig6} shows the dissipative dynamics of the XXZ model at the ferromagnetic critical point. As discussed previously, the trajectories of the zero-temperature and any finite-temperature initial states exhibit crossings, which is the hallmark of the sQME. Their overlap structures with the Liouvillian eigenmodes, however, differ significantly, as illustrated in the inset of Fig.~\ref{fig6}. More specifically, for finite-temperature states, the weight on the steady state increases with temperature, which is natural since the steady state corresponds to the infinite-temperature state. In addition, these states exhibit a noticeable nonzero distribution in the region of slow modes, which accounts for their intrinsically slow relaxation. By contrast, the zero-temperature state displays a qualitatively different overlap distribution: apart from the steady-state component, its nonzero overlaps are located at modes with large negative real parts of the eigenvalues (index of $\lambda_{\text{max}}$ is larger than $100$), while the overlaps with the slow modes vanish. This implies that its characteristic relaxation time is shorter than that of any finite-temperature state. This analysis explains the physical mechanism underlying the sQME between the zero-temperature and finite-temperature initial states: the absence of overlap with slow modes enables the zero-temperature state to relax faster despite being farther from equilibrium. At the same time, it also clarifies why no sQME occurs between different finite-temperature states, since their first nonzero overlaps lie in the same slow-mode region. As a consequence, the finite-temperature trajectories do not cross and remain nearly parallel, as observed in Fig.~\ref{fig6}.

The dissipative XXZ spin chain required for observing the sQME can be implemented on several state-of-the-art quantum simulation platforms. 
In ultracold atom experiments, XXZ-type Hamiltonians have been realized using optical lattices, where superexchange interactions provide tunable nearest- and next-nearest-neighbor couplings, and anisotropy can be controlled via interaction engineering \cite{XXZ_exp1,XXZ_exp2,XXZ_exp3,XXZ_exp4,XXZ_exp5,XXZ_exp6}. 
Dephasing-type dissipation, the key ingredient in our analysis, can be introduced through controlled noise, spontaneous photon scattering, or engineered coupling to tailored reservoirs, all of which have already been demonstrated experimentally. 
With these techniques, one can prepare initial thermal states at different effective temperatures, monitor their dissipative dynamics using quantum gas microscopy, and directly identify the crossing behavior that signals the presence of the sQME. 
Similar implementations are also feasible in trapped-ion chains and superconducting qubit arrays, where programmable interactions and tunable noise channels have been achieved. 
These advances ensure that the dissipative XXZ spin chain and the associated sQME can be experimentally realized and probed with existing technologies.

\section{Conclusion and outlook}
\label{Conclusion}

In this work, we have investigated the QME in open quantum spin chains subject to dephasing dissipation. By tracking the complete heating dynamics from thermal states at different initial temperatures, we identified the emergence of a sQME in the vicinity of quantum critical points. In this strong form, a state prepared farther from equilibrium relaxes faster than \emph{any} state initialized closer to the steady state--a marked departure from the wQME, which only requires a non-monotonic relaxation profile. By analyzing the spectral properties of the Liouvillian superoperator and the overlap distributions of different initial states with its eigenmodes, we established the physical mechanism underlying the emergence of sQME. In the XXZ chain, we demonstrated that the sQME arises precisely at the ferromagnetic ($\Delta=1$) and antiferromagnetic ($\Delta=-1$) quantum critical points, where the zero-temperature state relaxes faster than all finite-temperature states despite being farther from equilibrium. We further extended our analysis to the frustrated $J_{1}$–$J_{2}$ XXZ model, identifying a robust sQME in the parameter regime $J_{2}/J_{1}>-0.25$ at $\Delta=1$, and revealed its remarkable sensitivity to the frustration-induced phase transition at $J_{2}/J_{1}=-0.25$. Across all cases, we found that the sQME survives exclusively along the isotropic line $\Delta=1$, while deviations in anisotropy, system size, or temperature reduce it to a wQME. Our findings highlight the intimate connection between non-equilibrium relaxation dynamics and underlying quantum criticality. They also suggest that the QME provides a novel dynamical probe for detecting critical points and characterizing frustrated quantum phases. On the experimental side, the models we studied can be realized in ultracold atomic gases, trapped ions, and superconducting qubits, and dephasing-type dissipation has already been engineered in such platforms, making the observation of the QME experimentally feasible.

In future, several promising directions merit further exploration. First, it would be highly interesting to investigate whether the sQME persists in higher-spin chains ($S>1/2$), where richer on-site Hilbert spaces may lead to new relaxation pathways. Second, extending the analysis to higher-dimensional lattices could reveal whether dimensionality enhances or suppresses the effect, and whether novel universality classes of QME emerge. Third, studying the impact of long-range interactions, non-Markovian environments, and disorder would provide insight into the robustness of the phenomenon under realistic conditions. Finally, leveraging the QME as a tool for quantum state preparation or for accelerating thermalization in engineered reservoirs represents an intriguing avenue for applications in quantum simulation \cite{simulation,simulation1,simulation2} and quantum information processing. These directions could open a broad perspective for future theoretical and experimental studies of anomalous relaxation phenomena in open quantum systems.

\section*{Acknowledgements}
The work is supported by the National Key Re-
search and Development Program of China (Grants
No. 2023YFA1406704 and No. 2022YFA1405800) and
National Natural Science Foundation of China (Grant
No. 12304290 and No. 12474496). LP also acknowledges support from the Fundamental Research Funds for the Central Universities. \\


\end{document}